\documentclass[conference]{IEEEtran}
\IEEEoverridecommandlockouts

\usepackage{hyperref}
\usepackage{cite}
\usepackage{amsmath,amssymb,amsfonts}
\usepackage{algorithmic}
\usepackage{graphicx}
\usepackage{booktabs}
\usepackage{textcomp}
\usepackage{xcolor}
\usepackage{colortbl}
\usepackage{tabularx}
\def\BibTeX{{\rm B\kern-.05em{\sc i\kern-.025em b}\kern-.08em
T\kern-.1667em\lower.7ex\hbox{E}\kern-.125emX}}

\begin{document}

\title{``It Comes in Notebooks'': Changes and Challenges When
Operationalizing ML Prototypes}

\author{\IEEEauthorblockN{Arumoy Shome}
  \IEEEauthorblockA{\textit{Delft University of Technology}\\
    Delft, The Netherlands \\
  a.shome@tudelft.nl}
  \and
  \IEEEauthorblockN{Lu{\'\i}s Cruz}
  \IEEEauthorblockA{\textit{Delft University of Technology}\\
    Delft, The Netherlands \\
  l.cruz@tudelft.nl}
  \and
  \IEEEauthorblockN{Diomidis Spinellis}
  \IEEEauthorblockA{\textit{Delft University of Technology}\\
    Delft, The Netherlands \\
  d.spinellis@tudelft.nl}
  \and
  \IEEEauthorblockN{Arie van Deursen}
  \IEEEauthorblockA{\textit{Delft University of Technology}\\
    Delft, The Netherlands \\
  arie.vandeursen@tudelft.nl}
}

\maketitle

\begin{abstract}

  Machine learning practitioners commonly prototype models
  in computational notebooks
  before transitioning them to automated production systems.
  Despite its prevalence,
  the concrete engineering work involved in this transition
  and the software quality concerns that motivate it
  remain insufficiently characterized.
  We report on a qualitative study based on semi-structured interviews
  with 13 ML practitioners from industry and academia.
  Using reflexive thematic analysis,
  we identify 23 engineering changes
  organized into five themes:
  code restructuring,
  data pipeline development,
  testing \& validation,
  pipeline automation,
  and monitoring \& observability.
  We also identify 20 software quality attributes across
  the ML development lifecycle
  and map them to the engineering changes.
  A recurring pattern in our findings is that
  computational notebooks externalize oversight to the human practitioner,
  and defer costs that become obligatory at operationalization time.
  Operationalization constitutes the repayment of this technical debt
  accumulated during prototyping,
  which we refer to as \emph{oversight debt}.
  Practitioners do not merely restructure notebook code,
  but repay this debt by constructing automated substitutes
  for the interactive oversight that notebooks provide.
  We further present seven quality trade-offs
  showing that these tensions are properties
  of the notebook-to-production transition,
  rather than symptoms of poor engineering practice.
  Our findings structure operationalization effort,
  establish empirical links between engineering changes and software quality concerns,
  and provide implications for practitioners,
  tool designers,
  and researchers working on ML-enabled software systems.

\end{abstract}


\section{Introduction}

The development of ML-enabled software systems
commonly involves a transition from exploratory model development
to production-oriented software engineering.
In the early stages of the machine learning development lifecycle,
data scientists and other practitioners in model-development roles
focus on iterative data preparation and model experimentation,
through cycles of data cleaning,
feature engineering,
model training,
and evaluation~\cite{amershi2019software, haakman2021, martinez-plumed2021crisp-dm}.

Due to its nature, this work is often conducted
in computational notebooks, such as Jupyter notebooks~\cite{kluyverthomas2016}.
Despite prior studies showing challenges with software quality
and maintainability~\cite{pimentel2019large-scale, pimentel2021understanding, grotov2022large-scale, chattopadhyay2020what}, Jupyter notebooks continue to be the most widely used tool within the data science community~\cite{psallidas2019data, perkel2018why}.
This popularity stems from their support for interactive experimentation,
which allows code to be interspersed with textual explanations,
tables,
and visualizations.
However, the very properties that make notebooks effective for exploration,
such as interactivity,
non-linear execution,
and inline visualization,
also create friction
when the code must be operationalized into a reliable,
efficient,
and maintainable automated production system.

Once a suitable ML prototype is identified,
development shifts toward operationalizing it within a production environment,
which typically involves software engineers
or machine learning engineers~\cite{shankar2024we, nahar2023meta-summary}.
This shift requires
restructuring exploratory code into modular, testable components,
and additionally constructing entirely new technical capabilities
such as automated pipelines,
data validation checks,
and observability infrastructure,
to compensate for the loss of interactive oversight that notebooks provide.
We refer to these modifications collectively as
\textit{engineering changes},
encompassing modifications to source code,
infrastructure,
tooling,
and development practices
that practitioners perform as
part of operationalization.

Prior work has studied the challenges of building ML-enabled systems
from the perspective of ML artifact reuse~\cite{sens2024large-scale},
software development and maintainability~\cite{lwakatare2020large-scale},
deployment~\cite{chen2020comprehensive},
and collaboration within teams and organizations~\cite{nahar2022collaboration}.
However, these studies have primarily examined ML system challenges
either during the upstream developmental phase,
or in the final operations phase of the ML development lifecycle.
Consequently, the concrete engineering work
involved in transitioning from exploratory notebook artifacts to production-ready software
remain insufficiently characterized.
Furthermore, while software quality models
have been extended to account for the unique characteristics of ML systems~\cite{siebert2021construction},
there is limited empirical evidence
connecting specific operationalization practices
to the quality concerns that motivate them.

To address this gap,
we conduct semi-structured interviews with 13 machine learning practitioners
from academia and industry,
each with prior experience deploying machine learning models within production systems.
We use reflexive thematic analysis
to examine the collected data
and answer the following research questions.

\begin{description}
  \item[RQ1.] \textbf{What are the engineering changes
  that transition ML prototypes to production?}
  \item[RQ2.] \textbf{What are the software quality considerations
  when integrating ML prototypes into production?}
  \item[RQ3.] \textbf{What quality trade-offs arise
  from the loss of interactive oversight
  when operationalizing ML prototypes?}
\end{description}

The contributions of this study are threefold.
\textbf{First}, we provide a thematically organized catalog
of 23 engineering changes
that practitioners perform
when transitioning ML prototypes into production systems.
These changes are grounded in the experiences of practitioners
across academia and industry,
and address challenges identified in existing literature.
\textbf{Second}, we identify 20 software quality attributes that motivate these changes,
and establish an empirical link between the engineering work
of operationalization and software quality concerns.
\textbf{Third}, we find seven tensions and trade-offs
practitioners face due to the loss of a human
from the execution loop during this transition.
In such situations,
practitioners need to make deliberate engineering compromises
since addressing one quality concern can adversely affect another.

A recurring pattern across our findings
is that notebooks externalize oversight to the human practitioner,
and accumulate what we refer to as \emph{oversight debt}.
This debt is invisible until operationalization time,
and is repaid through the engineering changes that we identified.

Our findings offer researchers and practitioners
a structured understanding
of the engineering effort involved in moving ML prototypes to production,
and complements prior work
that has examined this transition
from the perspective of artifact reuse,
team collaboration,
and system maintainability.

\section{Study Design}

\begin{figure}
  \centering
  \includegraphics[width=\columnwidth]{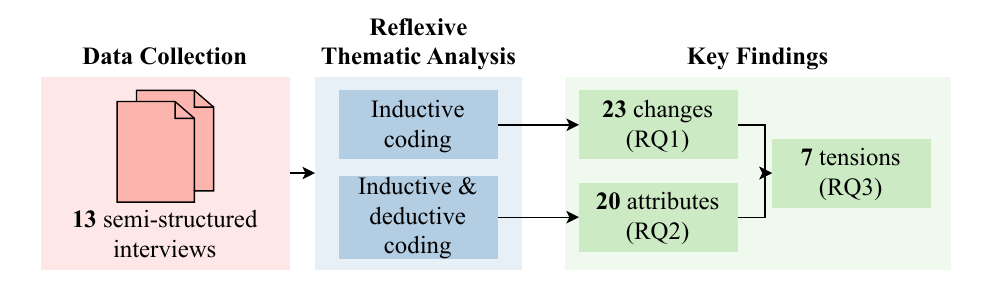}
  \caption{Summary of study design.}
  \label{tab:demographic}
  \label{fig:method}
\end{figure}

\begin{table}
  \centering
  \caption{Demographics}
  \begin{tabularx}{\columnwidth}{l X}
    \toprule
    \textbf{\emph{Company Characteristics}} & \\
    Size & Large (46\%), mid-size (7\%), small or startup (15\%),
    academia (30\%) \\
    Sector & Technology (54\%), consultancy (7\%), healthcare (15\%),
    banking and fintech (15\%), other (7\%) \\
    \midrule
    \textbf{\emph{Participant Characteristics}} & \\
    \midrule
    Role & Data scientists and engineers (+ data-, ML-,
    MLOps-engineers) (46\%), researcher (37\%), product manager
    (7\%), lead (7\%) \\
    \bottomrule
  \end{tabularx}
\end{table}

Using the Goal Question Metric Paradigm~\cite{caldiera1994goal},
the research objective of this study
is to \emph{characterize} the \emph{engineering changes}
involved in operationalizing ML prototypes,
from the perspective of \emph{software quality}.
This objective guides the data collection and analysis strategies
which are summarized in Fig.~\ref{fig:method} and described below.

\subsection{Data Collection}

\subsubsection{Participants}

We recruit ML practitioners with experience in deploying ML components
within production software systems.
Following the definition by Nahar et al.~\cite{nahar2025},
a \textit{production system} in this study
refers to a software product containing one or more ML components that serve
end-users (e.g., through a web or mobile application). Thus, our
inclusion criteria require that each participant has: (1)~professional
experience developing or deploying ML models, and (2)~involvement with at least
one ML-enabled software system that was used by end-users.

Participants are recruited through convenience
sampling~\cite{baltes2022}, leveraging the authors' professional
networks and social media outreach on \emph{X} and \emph{LinkedIn}. We actively seek
diversity across organizational contexts, professional roles, and
seniority levels. The resulting cohort (Table~\ref{tab:demographic})
spans participants from large enterprises (46\%), mid-size companies
(7\%), startups (15\%), and academia (30\%);
across technology, healthcare, and financial service sectors.
Roles range from data scientists and ML engineers
to product leads and academic researchers.

A total of 13 practitioners participated in the study, all based in
Europe, with a median of \textbf{4} years of professional experience
(range: \textbf{1--13} years).

\subsubsection{Interview Protocol}

A semi-structured interview guide is developed
in accordance with guidelines by Patton~\cite{patton2010qualitative},
through an iterative, collaborative process involving all authors.
The guide is organized around \textbf{4} thematic areas:
(1)~the participant's role and experience with ML systems,
(2)~the process of transitioning ML prototypes to production,
(3)~engineering changes made during operationalization,
and (4)~quality concerns and trade-offs encountered.
Follow-up probes are employed to explore specific topics in depth,
adapted to the direction of each interview.
The interview protocol is reviewed and approved by the university ethics board.

The interview guide is refined through two pilot interviews
conducted with PhD students from our research group.
For example, the pilots revealed that participants interpret
`deployment' differently.
This prompted us to add a clarifying preamble defining the term.
Pilot data is not included in the final analysis.

\subsubsection{Interview Execution}

13 interviews are conducted by the first author
between December 2024 and March 2025,
either in person (2) or via Microsoft Teams (11).
Each interview is 60--90 minutes long
and is conducted in English.
The first author guided the conversation using the interview guide
and took field notes during the sessions.
Interviews are audio-recorded with participant consent
and transcribed using the transcription tool built-in to Microsoft Teams.
The transcripts are later manually verified and corrected by the first author.

\subsection{Data Analysis}

We analyze the interview transcripts using reflexive thematic analysis
(RTA) following the six-step process described by Braun and
Clarke~\cite{braun2006, braun2019}.
In line with the RTA methodology,
codes reflect the first author's analytical interpretation of the data
rather than inter-rater agreement.
Methodological rigor is ensured
through ongoing dialogue with all authors
and iterative refinement of codes and themes~\cite{braun2019}.

Transcripts and codes are managed using spreadsheets
and the analysis is conducted in \textbf{two} complementary phases,
corresponding to RQ1 and RQ2.
Throughout both phases,
emerging codes and themes
are discussed with all authors in regular meetings.
New codes are reviewed,
similar codes are merged,
and disagreements are resolved through discussion
until consensus is reached.
This collaborative process
consolidated the initial 30 and 50 codes,
to the final set of \textbf{23} engineering changes (RQ1)
and \textbf{20} quality attributes (RQ2) respectively.

We observe thematic saturation in the last \textbf{two} interviews,
where existing codes were reinforced and no substantively new codes
emerged from the transcripts~\cite{aldiabat2018data}.

\subsubsection{Phase 1.\ Engineering Changes (RQ1)}

For RQ1, we employ a purely inductive coding approach.
The first author read all transcripts in multiple passes
to develop familiarity with the data (\emph{step 1: familiarization}).
Segments describing engineering modifications,
such as changes to code,
infrastructure,
tooling,
or development practices made during operationalization,
are assigned initial codes (\emph{step 2: generating initial codes}).
Through iterative refinement,
codes describing similar changes are consolidated
to yield a set of 23 distinct codes (\emph{step 3: searching for themes}).
For instance, codes `adding data documentation'
and `migrating markdown cells to documentation system'
were merged into `EC4.\ Decision documentation migration'.
These codes are then organized into five higher-level themes
based on the nature of the engineering work involved (\emph{step 4: reviewing themes}).
Theme definitions are refined through discussion
with all authors until consensus is reached (\emph{step 5: defining and naming themes}),
resulting in the catalog reported in Sec.~\ref{sec:results}.

\subsubsection{Phase 2.\ Quality Considerations (RQ2)}

For RQ2, we employ a hybrid deductive-inductive approach.
We use the ML system quality model proposed by Siebert et al.~\cite{siebert2021construction}
as an initial coding framework
which is constructed using the ISO/IEC 25010 standard~\cite{iso25002}
and prior literature
that focus on software quality
for ML systems~\cite{fujii2020guidelines, marselis2018machine,
  belani2019requirements,
bosch2018takes, horkoff2019non-functional, ismail2019}.

Transcript segments expressing quality concerns
are first coded against the attributes in Siebert et al.'s quality
model (\textit{deductive coding}).
When a segment describing a quality concern is not captured by this model,
the first author followed the attribute elicitation methodology
proposed by Siebert et al.~\cite{siebert2021construction}
(\textit{inductive coding}).
The first author familiarized themselves with the use case and
application context,
and identified the relevant ML quality requirements from the transcripts.
For each segment, we then identify the relevant \textit{entity}
and characterize the \textit{quality attribute} of concern.
For example, the quality attribute `correctness'
can be associated with the `development data'
and the `trained model' entities.
The former
is concerned with the quality of the training data
(e.g., the degree to which it is free from errors such as missing values)
while the latter
is concerned with the correctness of the predictions of the model
(e.g., accuracy, precision and recall).

\section{Results}
\label{sec:results}

\begin{table*}
  \centering
  \caption{Summary of engineering changes, quality attributes and challenges
    across the five themes identified in our analysis. Quality
    attributes are marked as improved~($\uparrow$) by the engineering
    changes within each theme.
    Challenges describe quality trade-offs~($\leftrightarrows$),
    showing which attributes are improved~($\uparrow$) at the expense
  of others~($\downarrow$).}
  \label{tab:summary}
  \begin{tabularx}{\textwidth}{p{3cm} X p{2.6cm} X}
    \toprule
    \textbf{Theme \& Participant Coverage} &
    \textbf{Engineering Changes} &
    \textbf{Improved Quality Attributes} &
    \textbf{Challenges \& Trade-offs} \\
    \midrule
    \textbf{\emph{Code Restructuring}} \newline
    {\footnotesize (Sec.~\ref{sec:restructuring})} \newline
    {\footnotesize P1, P3, P4, P5, P6, P7, P12, P13} &
    \textbf{EC1.}~Exploratory artifact removal \newline
    \textbf{EC2.}~Code modularization \newline
    \textbf{EC3.}~Data source abstraction \newline
    \textbf{EC4.}~Decision documentation migration \newline
    \textbf{EC5.}~Codebase organization \newline
    \textbf{EC6.}~Hardware agnostic adaptation &
    Maintainability \newline
    Reusability \newline
    Understandability \newline
    Traceability \newline
    Portability \newline
    Reliability &
    \textbf{CT1.}~Restructuring depth vs.\ prototyping velocity \newline
    {\footnotesize $\uparrow$~Maintainability $\leftrightarrows$
    $\downarrow$~Dev.\ velocity}\\
    \midrule
    \textbf{\emph{Data pipeline development}} \newline
    {\footnotesize (Sec.~\ref{sec:datapipeline})} \newline
    {\footnotesize P1, P2, P3, P6, P8, P9, P11, P12} &
    \textbf{EC7.}~Data processing scalability \newline
    \textbf{EC8.}~Feedback loop mitigation \newline
    \textbf{EC9.}~Intermediate data snapshotting &
    Scalability \newline
    Data integrity \newline
    Reliability \newline
    Debuggability &
    \textbf{CT2.}~Dev--production environment divergence \newline
    {\footnotesize $\uparrow$~Dev.\ velocity $\leftrightarrows$
    $\downarrow$~Portability} \\
    \midrule
    \textbf{\emph{Testing \& validation}} \newline
    {\footnotesize (Sec.~\ref{sec:testing})} \newline
    {\footnotesize P1, P3, P4, P7, P8, P9, P12, P13} &
    \textbf{EC10.}~Data quality assertions \newline
    \textbf{EC11.}~Mock data generation \newline
    \textbf{EC12.}~Pre-deployment drift testing \newline
    \textbf{EC13.}~Fail-safe pipeline checks \newline
    \textbf{EC14.}~Production validation (A/B testing) &
    Data quality \newline
    Model correctness \newline
    Testability \newline
    Reliability &
    \textbf{CT3.}~Testing investment vs.\ technology change \newline
    {\footnotesize $\uparrow$~Reliability $\leftrightarrows$
    $\downarrow$~Dev.\ velocity}\\
    \midrule
    \textbf{\emph{Pipeline automation}} \newline
    {\footnotesize (Sec.~\ref{sec:pipeline})} \newline
    {\footnotesize P1, P2, P3, P5, P6, P7, P8, P12} &
    \textbf{EC15.}~Script migration \newline
    \textbf{EC16.}~ML framework adoption \newline
    \textbf{EC17.}~Dependency mgmt.\ \& containerization \newline
    \textbf{EC18.}~Pipeline guardrails \newline
    \textbf{EC19.}~Technology stack adaptation &
    Maintainability \newline
    Portability \newline
    Reusability \newline
    Reproducibility \newline
    Cost efficiency \newline
    Reliability \newline
    Compatibility &
    \textbf{CT4.}~Legacy system integration \newline
    {\footnotesize $\uparrow$~Reliability $\leftrightarrows$
    $\downarrow$~Dev. velocity} \newline
    \textbf{CT5.}~Notebook--script duality \newline
    {\footnotesize $\uparrow$~Maintainability $\leftrightarrows$
    $\downarrow$~Interactivity}\\
    \midrule
    \textbf{\emph{Monitoring \& Observability}} \newline
    {\footnotesize (Sec.~\ref{sec:monitoring})} \newline
    {\footnotesize P1, P3, P4, P5, P6, P7, P8, P9, P10, P12} &
    \textbf{EC20.}~Structured logging \newline
    \textbf{EC21.}~Experiment tracking \newline
    \textbf{EC22.}~Prediction monitoring \& alerting \newline
    \textbf{EC23.}~Audit trail construction &
    Debuggability \newline
    Reproducibility \newline
    Runtime correctness \newline
    Traceability &
    \textbf{CT6.}~Privacy constraints on monitoring \newline
    {\footnotesize $\uparrow$~Privacy $\leftrightarrows$
    $\downarrow$~Runtime correctness} \newline
    \textbf{CT7.}~Organizational silos \& silent failures \newline
    {\footnotesize $\downarrow$~Comm.\ overhead $\leftrightarrows$
    $\downarrow$~Runtime correctness}\\
    \bottomrule
  \end{tabularx}
\end{table*}

This section is organized by the five themes that emerged from the data.
For each theme,
we present the engineering changes~(\textbf{EC}) practitioners
perform during operationalization (RQ1),
the software quality attributes that motivate them (RQ2),
and reflect on how these concerns are intertwined in practice.
We also present the challenges and trade-offs~(\textbf{CT})
in situations where addressing one quality concern
conflicts with another,
forcing practitioners to make deliberate engineering compromises (RQ3).
Table~\ref{tab:summary} provides an overview of the 5 themes,
23 engineering changes,
20 quality attributes,
and 7 challenges identified in this study.

\subsection{Code Restucturing}
\label{sec:restructuring}

Notebook code is reorganized into modular, well-documented, and hardware-agnostic
production artifacts to improve maintainability, reusability and understandability.

\textbf{EC1. Exploratory artifact removal:}
Notebooks developed during prototyping contain exploratory artifacts,
such as inline visualizations,
intermediate result tables,
and abandoned experimental
branches,
which are useful during model development
but have no role
in a production pipeline.
To reduce the surface area of the codebase to only relevant logic, practitioners
(P4, P5, P7) eliminate exploratory artifacts that can otherwise create
confusion for future maintainers about which components are essential and which
are residual.

\textbf{EC2. Code modularization:}
Practitioners restructure the linear,
procedural scripting style of notebook code
into functions, classes, and modules.
Modular code can be reasoned about in isolation,
reused across pipeline stages,
and tested independently (as described in Sec.~\ref{sec:testing}).
P1 noted how this transformation improved collaboration,
making it easier for new team members
to comprehend the code: ``they can look at these eight
lines of code and say, \emph{okay, I understand}''.
Notably, P13 reported that anticipating the eventual migration to scripts
influenced early prototyping decisions.
P13 wrote modular code within the notebook ``with classes and functions''
from the outset,
which made the subsequent transition to automated scripts ``easier''.

\textbf{EC3. Data source abstraction:}
Production systems typically ingest data from multiple heterogeneous sources.
Practitioners implement importer classes and methods
to represent different data sources
such as ``import from Kafka'',
``import from Elastic Search''
and ``import from CSV''.
Encapsulating source-specific ingestion logic behind a common interface,
allows practitioners to add new data sources
without modifying the downstream pipeline code.
The abstraction also decouples the ML component from any single data source,
and improves reusability, maintainability and portability.

\textbf{EC4. Decision documentation migration:}
% \textit{traceability} and \textit{understandability}
Practitioners document technical decisions
and implementation rationale
in the markdown cells of notebooks during prototyping.
When code is migrated to scripts,
this documentation is transferred to
external technical documentation systems
to preserve it for future reference.
Practitioners (P3, P4, P5) emphasized that it is important to maintain this knowledge
``so that at the end we can refer back to it---why did we choose this model,
why did we go in this direction,
why did we filter this out---so we can actually justify our decisions''~\textsubscript{P4}.
Additionally, P5 recommended documenting data semantics
such as ``what each of the columns means and how I gathered these columns''.
This documentation facilitates onboarding new team members and external collaborators,
supports formal audits in regulated domains such as banking,
and helps maintain institutional memory across team changes.

\textbf{EC5. Codebase organization:}
% \textit{maintainability} and \textit{reusability}:
After modularizing individual components,
practitioners establish consistent codebase organization conventions
to reduce the cognitive overhead of navigating the project
and create clear boundaries for code ownership,
parallel development,
and code reuse across pipelines.
P6 described structuring code into a ``lib'' directory
containing reusable functions
and a ``pipeline'' directory
containing pipeline definitions
that use the functions defined in lib.
This separation enables tracking experiments
using branch-based development practices.
P6 described creating a ``new branch'' for each experiment,
making changes in the lib directory,
and triggering the pipeline to observe the effects.
At a higher level,
P12 described organizing different models into their own repositories,
so that practitioners working on a new feature
can ``fork'' the repository and start working independently.

\textbf{EC6. Hardware agnostic adaptation:}
% \textit{portability} and \textit{reliability}:
Production infrastructure can vary across deployment environments,
and hardcoding hardware assumptions
can create fragile deployments
that fail when the expected resources are unavailable.
Practitioners adapt inference code
to execute across heterogeneous hardware configurations,
by writing code that gracefully downgrades inference
from GPU to CPU when GPU resources are not present.

\textbf{CT1: Restructuring depth versus prototyping velocity:}
The practices that maximize exploration speed
also incur the highest restructuring cost during operationalization.
Notebook coding conventions create structural liability during operationalization,
as linear scripts must be decomposed into callable, reusable modules (EC2) with well-defined interfaces;
inline visualizations must be replaced (EC1) with headless logging and monitoring hooks;
hardcoded paths must be parameterized through configuration files (EC5) or environment variables;
and implicit state must be made explicit through well-defined data flow.
Front-loading this restructuring work during prototyping
is only rational if the practitioner knows the prototype will eventually be deployed,
in what form,
and under what constraints.
In practice however,
many prototypes are abandoned, pivoted, or repurposed in ways that were not anticipated
due to the lack of such knowledge that is inherent during genuine exploration.

\subsection{Data Pipeline Development}
\label{sec:datapipeline}

Data processing pipelines are scaled to production volumes
and data snapshots are added to improve reliability and debuggability.

\textbf{EC7. Data processing scalability:}
ML prototypes are typically developed using a sample of data
stored locally as a static CSV file.
Once operationalized,
practitioners reason about computational complexity
and resource constraints
as training is scaled
to the full production dataset with a ``much larger volume''.
As P1 noted, ``even a simple computation
like taking the average over one column can take a very long time''.
Practitioners described three techniques
targeting different components of the ML pipeline to address the scalability requirements.
Practitioners (P1, P3, P11) migrated data operations
from local, single-machine libraries to distributed computing frameworks.
P1 adapted data transformation operations written in Pandas~\cite{pandas}
to the API and execution model of Spark~\cite{spark}:
``so instead of making the computation on a pandas DataFrame,
we use a Spark DataFrame''.
P2 adapted the training algorithms themselves
by ``chunking the data''
and feeding it to the model in smaller batches,
in memory-constrained compute environments.
P9 described optimization work at the database level
such as adding ``database indexes'',
eliminating ``cross joins'',
and replacing ``inefficient'' SQL queries
to reduce data retrieval latency.

\textbf{EC8. Feedback loop mitigation:}
Sculley et al.~\cite{sculley2015} identified that
deployed models can inadvertently influence their own future training data,
through feedback loops that gradually degrade prediction quality
and accumulate technical debt.
This occurs when predictions on unseen production data
are fed back into the pipeline for retraining
which reinforces the model's biases over successive training cycles.
P6 described a system-level intervention to mitigate this risk
by tagging the portion of their internet traffic that is not served by any ML model.
This model-independent data is then used for retraining
to ensure that ``[our] data is not biased by models already in production''.

\textbf{EC9. Intermediate data snapshotting:}
To improve the reliability and debuggability of data pipelines,
practitioners introduce checkpoints that persist intermediate data pipeline outputs.
P1 described this practice: ``often we want to save intermediate steps as well
for debugging purposes or in case of crashes,
so the process can be resumed from that point rather than from the beginning.''
This change is particularly important for long-running data pipelines
where reprocessing from the beginning after a failure
incurs significant computational cost.
Data snapshots also allow practitioners to inspect the state of data at each pipeline stage,
compensating for the loss of interactive, cell-by-cell inspection that notebooks provide.

\textbf{CT2. Development--production environment divergence:}
Practitioners faced version and dependency conflicts
between their local development environments
and the distributed infrastructure used in production (EC7).
Local environments offer faster development velocity,
but cannot fully replicate the configuration of production systems,
particularly for distributed frameworks such as Spark
that introduce their own runtime dependencies.
P1 described this friction: ``since Spark is running somewhere else,
it means it could have a different version of Python
than what you're writing code in''.
This divergence means that code validated locally
may fail when executed on production infrastructure
and undermine confidence in the development workflow.

\subsection{Testing and Validation}
\label{sec:testing}

Automated checks and validations are constructed to improve data quality,
model correctness and testability.

\textbf{EC10. Data quality assertions:}
Practitioners introduce automated tests
to enforce data quality expectations that hold across pipeline executions.
Visual inspections conducted manually in notebooks
are replaced with unit tests to validate properties
such as value ranges,
presence of expected features,
and missing data thresholds (P4, P9, P12).
As P4 summarized, beyond a certain threshold of missing values,
the data is simply ``not of high quality''.
Integration tests complement these checks
by verifying model behavior under production conditions.
P12 described how code modularization (Sec.~\ref{sec:restructuring})
enabled fine-grained unit testing.
Since different business domains (e.g., flights, hotels, car rentals)
are represented as distinct classes in their code base,
adding a new feature could be tested by verifying that
``the number of features increased by one''.

\textbf{EC11. Mock data generation:}
To enable unit testing,
practitioners develop mock data generators
that simulate the characteristics of production data (P1, P12).
Unit tests are executed against this synthetic data,
while integration tests use real data in dedicated test environments:
``engineers deploy to a test environment where they would generate the data''~\textsubscript{P12}.
However, practitioners acknowledge that
it is difficult to ``truly test ML components'' using synthetic data
which cannot replicate the scale and distributional properties of production data.

\textbf{EC12. Pre-deployment drift testing:}
To address the limitations of testing using mock data
and to validate model robustness before deployment,
P3 and P7 described testing models using data samples
with ``week-by-week'' shifts to simulate data drift observed in production.
P7 conducted testing incrementally,
starting with a smaller geographic subset (e.g., popular regions first)
and progressively scaling to an entire country and eventually a continent.
P8 focused on edge-case identification
and created targeted data subsets
``by filtering by policy level,
by claimed amount,
by disease'' to identify inputs that the model could not handle reliably.

\textbf{EC13. Fail-safe pipeline checks:}
Beyond testing model outputs,
practitioners introduce checks that halt pipeline execution
when data quality violations can compromise downstream business decisions.
P9 described implementing unit tests that intentionally
``break the pipeline'' when quality thresholds are violated:
``I don't want to make this data available in a reporting interface.
If this happens, don't compute this data. It needs to be solved''.
Unlike data quality assertions that flag issues for review,
fail-safe checks enforce a hard constraint
preventing the pipeline to proceed until the violation is resolved.

\textbf{EC14. Production validation through A/B testing:}
Despite pre-deployment testing efforts,
practitioners consistently report that there is no substitute for validating model correctness in production.
P12 used ``A/B tests'' and ``interleaving experiments''
to evaluate how models perform against real user behavior,
production-scale data volumes,
and live distributional characteristics.
This final validation layer ensures model correctness
by comparing model variants under authentic conditions
that pre-deployment testing cannot fully replicate.

\textbf{CT3. Testing investment versus technological change:}
Ensuring data quality and model correctness requires substantial engineering effort,
such as writing assertions,
building mock data generators,
designing drift simulations
and configuring A/B testing infrastructure (EC10, EC11, EC12, EC14).
However, practitioners noted that the ML technology landscape evolves rapidly
as significant shifts occur ``every three or four years''.
This pace of change raises a sustainability concern
since testing infrastructure built around a specific framework or model architecture
may require significant rework as the underlying technology evolves.
While investing in testing infrastructure improves the current system reliability,
it also creates long-term maintenance obligations that can impede adoption of newer technologies.

\subsection{Pipeline Automation}
\label{sec:pipeline}

Notebook code is migrated into executable, automated pipelines to improve
maintainability, reproducibility and portability.

\textbf{EC15. Script migration:}
All practitioners use computational notebooks to develop initial ML prototypes
and subsequently migrate the relevant code into Python scripts for production use.
This migration is described as a prerequisite for adopting software engineering practices
that notebooks do not natively support.
Practitioners (P1, P3, P5, P6, P7, P8, P12) report that script-based development
enabled version control through ``Git'',
collaborative ``code reviews'',
and automated deployments through ``CI/CD'' pipelines,
that improve maintainability and source code provenance.
P6 described how the transition further facilitates scaling experiments to cloud computing infrastructure,
enabling different teams within the organization
to reuse trained models through shared pipeline artifacts.

\textbf{EC16. ML framework adoption:}
To impose structure on the migrated code,
practitioners adopt ML frameworks that enforce modular organization.
Frameworks such as Kedro~\cite{kedro} and Amazon SageMaker~\cite{liberty2020} are used
to decompose pipelines into distinct stages
of ``data ingestion'', ``model training'' and ``evaluation''.
P3 noted that Kedro enforces code to be organized into discrete ``nodes'' and ``pipelines'',
represented using functions and classes respectively.
This encapsulation improves reusability
since the same ``predict'' function
can be applied to a sample of training data during development
or to unseen data in production.

\textbf{EC17. Dependency management and containerization:}
% portability; reliability
Practitioners adopt dependency management and containerization practices
to ensure pipelines execute consistently
across heterogeneous computing environments
such as personal laptops and cloud infrastructure.
P1 recommended tracking ``the specific version of Python''
and all libraries with ``the specific versions''
to ensure compatibility and ``be reproducible on different machines''.
However, dependency management in ML projects remains an ongoing challenge.
External libraries can introduce ``breaking changes'' during development
and make dependencies incompatible with one another (P1, P3).
The move to automated pipelines
enables practitioners to detect such conflicts early through CI/CD integration,
as opposed to discovering them ``a year later'',
thereby ``allowing you to plan workarounds''~\textsubscript{P1}.

\textbf{EC18. Pipeline guardrails:}
% cost efficiency; reliability
Practitioners introduce automated checks
to prevent costly failures during pipeline execution.
Unlike the interactive notebook environment where practitioners visually inspect intermediate outputs,
automated pipelines require programmatic safeguards that substitute for human oversight.
At the training level,
P5 used early stopping callbacks in the Hugging Face Transformers library~\cite{wolf2020}
to halt training when ``the loss doesn't decrease for like three epochs in a row''
to prevent unnecessary computation.
At the pipeline level,
P6 performed validation checks before execution
to ``avoid errors before spending money'',
by implementing assertions to verify date ranges,
dataset consistency,
and feature availability.
Assertions are also used to
enforce model complexity constraints
that reduce inference latency,
by restricting training to ``shallow trees so the models aren't heavy''~\textsubscript{P6}.

\textbf{EC19. Technology stack adaptation:}
% compatibility
The most suitable algorithm for prototyping
may not match the frameworks and tools supported in the production environment.
P6 described such an instance
where the ML prototype had to be re-implemented
to align with the organization's existing infrastructure and technology choices.
The prototype was obtained from an academic paper
and ``was originally implemented in TensorFlow'',
but had to be re-implemented using ``LightGBM''
to match the production stack.

\textbf{CT4. Legacy system integration:}
While newer projects can adopt automated pipelines from their inception,
P1 and P12 mentioned integrating modern tooling into legacy systems
as a source of friction
that requires manual intervention and ``coordinated effort across stakeholders''.
In particular, script migration (EC15) and technology stack adaptation (EC19)
become significantly costlier
when the target environment is constrained by existing infrastructure.
Development velocity for ML components in legacy systems
is slower due to their technical complexity
and because these components are often ``critical to the success'' of the business.
To ensure such revenue-critical components continue to function reliably,
all changes must be thoroughly vetted using ``A/B testing''
and extensive ``user studies''
to ``make sure you don't break stuff''~\textsubscript{P12}.

\textbf{CT5. Notebook--script duality:}
Although practitioners acknowledge
that notebooks become ``cumbersome'' for managing multiple experiments,
the transition to scripts (EC15) introduced a persistent tension in their workflow.
Debugging is perceived as ``more intuitive'' in notebooks,
which support inspection of intermediate data through tables and visualizations.
This interactivity is however lost in traditional IDEs and debuggers,
which are optimized for text-based inspection,
making it ``not easy to display'' data tables.
P3 and P5 described this as a ``frustrating'' experience of maintaining ``two versions of the same file'',
one in notebooks for interactive exploration,
and another in a Python script for production execution.

\subsection{Monitoring and Observability}
\label{sec:monitoring}

Instrumentation is added to gain visibility into model and pipeline behavior to improve
debuggability, traceability and runtime correctness.

\textbf{EC20. Structured logging:}
Practitioners rely on print statements and cell outputs
to inspect intermediate results during the linear execution of code cells in notebooks.
This approach becomes impractical in automated pipelines,
where multiple components execute asynchronously on external infrastructure.
To diagnose failures in distributed environments,
practitioners (P1, P5, P7, P9, P10) incorporate logging statements into their training and inference code.
The logs are stored in databases
and consumed by external observability platforms such as Grafana~\cite{grafana} and Tableau~\cite{tableau},
which provide dashboards for analyzing pipeline behavior during training and inference.

\textbf{EC21. Experiment tracking:}
It is ``much harder to keep track of experiments'' in notebooks,
where results are often scattered across code cells and multiple file versions.
To address this, practitioners integrate experiment tracking frameworks
such as Weights \& Biases~\cite{wandb} into their automated pipelines (P3, P6, P8).
These tools automatically log
``model performance metrics,'' ``feature importance,'' and ``loss curves'',
producing reports that enable comparison across experimental runs.
P6 noted that this integration improved both reproducibility and traceability,
since every experiment is recorded with its parameters and outcomes
thus providing a persistent record of decisions made during model development.

\textbf{EC22. Prediction monitoring and alerting:}
Once models are deployed,
practitioners use a combination of automated alerting and manual inspection
to detect degradation in model behavior during runtime.
Practitioners (P3, P7, P8) use observability platforms
to visualize model predictions over time,
and configure automated alerts to notify engineers of anomalous behavior.
For instance, P8 described alerts that triggered when ``fewer claims are being auto-approved'',
signaling a potential increase in fraudulent activity.
For P3, automated alerts are escalated to engineers
whenever ``missing data'' is encountered in production data feeds.
However, practitioners emphasized that automated monitoring alone
is insufficient to assess runtime correctness.
Manual review of prediction trends remains necessary,
typically by comparing current outputs against ``the previous several weeks''
to identify gradual model degradation.

\textbf{EC23. Audit trail construction:}
To support traceability and post-hoc analysis,
practitioners track input data alongside runtime predictions,
and create end-to-end audit trails
that link model outputs to their inputs and pipeline configurations.
P10 described maintaining detailed records of ``which data inputs produced which predictions''
that enabled the team to investigate anomalous outputs retroactively.

\textbf{CT6. Privacy constraints on monitoring:}
Prediction monitoring and audit trail infrastructure (EC22, EC23)
are necessary to ensure runtime correctness,
but not always feasible due to data privacy and security requirements.
P4 described such an instance: ``like when we are handling sensitive client data,
it's fundamental that we don't store any information''.
In such settings, practitioners are forced to rely on weaker, indirect signals
such as ``user feedback'' to monitor model health.

\textbf{CT7. Organizational silos and silent failures:}
Communication gaps between development and operation teams
can limit the monitoring effectiveness that structured logging and alerting infrastructure provide (EC20, EC22).
Reducing cross-team coordination can cut communication overhead and save time,
but increases the risk of undetected failures that can cause significant financial harm.
P7 recalled a situation
where development decisions were deliberately not communicated to the operations team
to minimize coordination overhead.
This however created conditions for silent data errors
that did not explicitly break the pipeline
but gradually degraded model correctness
and propagated into business decisions.
P7 recounted how a modification to an
underlying SQL query in the data pipeline
could ``have been disastrous\ldots close to millions of euros in damages''.
Following this incident, automated alerts were introduced to detect such discrepancies immediately.

\section{Discussion}

Our findings show that operationalizing ML prototypes
is not a straightforward migration of code from notebooks to scripts,
but involves constructing entirely new technical capabilities
to compensate for the loss of interactive, human-in-the-loop workflows.
In this section, we interpret our key findings,
connect them to prior work,
and discuss implications for researchers, practitioners, and tool designers.

\subsection{Operationalization as Repayment of Technical Debt}

A recurring pattern across the 23 engineering changes we identified
is that they programmatically replicate capabilities that notebooks provide interactively.
In notebooks, practitioners visually inspect intermediate data outputs,
manually verify data quality through inline tables,
and iteratively debug by re-executing individual cells.
When code moves to automated pipelines,
these interactive safeguards disappear
and practitioners must construct automated substitutes.
We interpret this pattern as the repayment of a form of technical debt,
that we call \textit{oversight debt}.

Ward Cunningham introduced the concept of technical debt
and Sculley et al.~\cite{sculley2015} subsequently extended it to ML systems.
Technical debt can be used to explain engineering obligations
that are deferred during development,
but must be repaid later with interest.
Sculley et al.\ identified debt such as
undeclared system boundaries,
entangled data dependencies,
and feedback loops
that accumulate at the system architecture level.
Our findings reveal
a complementary form of debt
that is not introduced by poor architectural decisions,
but by the computational notebook format itself.

Notebooks externalize oversight to the human practitioner at execution time.
Every cell execution is an implicit checkpoint,
that is used by the practitioner
to observe intermediate outputs,
judge the data quality,
and decide whether to proceed.
This oversight is not encoded in the artifact,
but exists in the practitioner's cognitive loop.
The prototyping phase is therefore not debt-free,
even when the notebook code is clean and well-structured.
Rather, it defers the cost of oversight to a human operating interactively,
which makes the debt invisible until the human is removed from the loop.

Operationalization is the moment when this debt becomes due.
The engineering changes our participants described are, in large part, the repayment mechanism.\
Intermediate data snapshotting (EC9, Sec.~\ref{sec:datapipeline}) repays the debt
of cell-by-cell execution as the mechanism for
inspecting data state across pipeline stages.
Data quality assertions (EC10, Sec.~\ref{sec:testing}) repay the debt
of visual, ad-hoc inspection of data tables.
Pipeline guardrails (EC18, Sec.~\ref{sec:pipeline}) repay the debt
of human judgment exercised interactively
when deciding whether to proceed with a computation.
Structured logging (EC20, Sec.~\ref{sec:monitoring}) repays the debt
of relying on inline cell outputs to inspect pipeline behavior.

This framing has important implications
for how the engineering cost of operationalization should be understood.
The conventional narrative is
that notebook code is messy
and that operationalization involves cleaning it up,
by improving structure,
enforcing modularity,
and adopting better practices.
This characterization is only partially correct.
The code restructuring changes in Sec.~\ref{sec:restructuring}
do address quality deficiencies that notebooks encourage.
However, it does not account for a substantial portion of the engineering work we observed.
Many of the changes our participants described
are not corrections of deficient code
but obligatory repayments of oversight debt.

This distinction carries a practical consequence for project planning.
Unlike conventional technical debt,
which is partially avoidable through better upfront practices,
oversight debt is structurally unavoidable.
It is incurred by virtue of using the notebook format for prototyping
and is only repayable at operationalization time.
Organizations that budget operationalization effort solely around code restructuring and refactoring,
risk underestimating the work involved,
because the compensatory infrastructure required to repay oversight debt
often represents the larger and less predictable investment.

\subsection{Quality Trade-offs as Structural Tensions}

Our analysis identified seven challenges and trade-offs that practitioners navigate during operationalization.
We find that these trade-offs are not symptoms of poor engineering practice,
rather \textit{structural tensions} inherent to the notebook-to-production transition,
where two legitimate quality goals are in genuine conflict
and improving one comes at the cost of the other.

The most pervasive tension is between
\textit{maintainability}, a software quality attribute,
and \textit{development velocity}, which is associated with the development process
rather than the quality of the software.
The practices that make prototyping fast,
such as writing linear scripts,
using inline visualizations,
and hardcoding data paths,
are also those that create the highest restructuring cost when the code must go to production (CT1, Sec.~\ref{sec:restructuring}).
P13 demonstrated that writing modular code in notebooks from the outset can reduce this cost,
but this front-loads effort during a phase where it is inherently uncertain
whether the prototype will be selected for deployment.
If a data scientist invests in clean,
modular notebook code for a prototype that is ultimately discarded,
that effort is wasted.
The trade-off is irreducible since practitioners either pay earlier
through slower exploration
or pay later through harder operationalization.
Although neither choice is wrong,
the appropriate balance depends on the likelihood that the prototype will ship.

Other trade-offs involve tensions between \textit{team-level} and \textit{system-level} quality concerns.
The organizational silos challenge (CT7, Sec.~\ref{sec:monitoring})
exemplified that reducing communication overhead between development and operations teams
improved team-level efficiency
but created the conditions for silent data errors that nearly caused millions of euros in damages.
This challenge echoes the collaboration challenges reported by Nahar et al.~\cite{nahar2022collaboration}
and adds the \textit{engineering change} dimension.
We show not only that cross-team collaboration is challenging,
but what specific engineering work (such as structured logging and automated alerting)
practitioners perform to mitigate these collaboration failures.

Finally, some trade-offs present tensions with no clean engineering resolution.
In regulated domains, the monitoring infrastructure needed to ensure
runtime correctness directly conflicts with data protection obligations
(CT6, Sec.~\ref{sec:monitoring}).
Similarly,
organizations must find a sustainable testing infrastructure in
the rapidly changing ML technology landscape which shifts
significantly every few years (CT3, Sec.~\ref{sec:testing}).
Organizations face a choice between comprehensive but potentially short-lived testing
infrastructure and lighter-weight tests that provide weaker guarantees
but are easier to replace.

Collectively, these trade-offs indicate that
operationalization is not a one-time migration with a defined endpoint.
The tensions we identified persist throughout the system's lifecycle as
requirements change, teams evolve, and technologies shift. Organizations
should therefore view operationalization not as a project with a
completion date, but as an ongoing engineering activity that requires
sustained investment and deliberate decision-making about which quality
attributes to prioritize in which contexts.

\subsection{The Dual Lifecycle Challenge}

The notebook--script duality (CT5, Sec.~\ref{sec:pipeline}) identified in our results
is analogous to the synchronization problem of model-driven software development (MDSD).
In MDSD, a high-level model,
such as a UML diagram or a domain-specific model,
serves as the primary development artifact from which executable code is generated~\cite{schmidt2006model}.
Round-trip engineering~\cite{angyal2008synchronizing} addresses the challenges that arise when changes flow in both directions
i.e., forward engineering propagates model changes into code,
while reverse engineering propagates code changes back into the model.
The two representations can drift apart when both artifacts are modified independently
and changes are not consistently propagated in both directions.

The notebook--script duality exhibits the same structure.
When practitioners maintain both the notebook for interactive debugging
and the script for automated execution,
changes to either artifact may not be propagated to the other.
Subtle inconsistencies in data preprocessing,
feature engineering,
or model configuration
can then emerge between the two artifacts
and silently alter model predictions without raising explicit errors.
Unlike MDSD however,
there is no automated tooling to sync the changes between notebooks and scripts,
making this divergence difficult to detect and potentially costly to diagnose.

This risk is further compounded by the executable nature of the notebook.
Unlike a UML model,
the notebook may continue to serve as an active environment for experimentation and debugging
even after the production script is deployed.
Practitioners may update the notebook during ongoing investigation,
introduce changes that are never propagated to the script,
or rely on notebook outputs to reason about a production system that no longer behaves identically.

We advise practitioners navigating this challenge to consider migrating
to modern tools such as nbdev~\cite{nbdev} and Kedro~\cite{kedro}
that eliminate managing two artifacts indefinitely.
The engineering implications and trade-offs of adopting such tools
are discussed in the next section.

\subsection{Implications for Tool Design}
\label{sec:implications-tools}

Several participants (P3, P5, P8) reported frustrations with traditional debugging interfaces when working with ML code.
P5 observed that inspecting data tables is ``not easy to display'' in conventional IDEs,
which are optimized for text-based inspection
and lack native support for tabular and visual data.
This points to a broader need for development environments
that extend traditional IDEs with the interactive data inspection capabilities that notebooks provide.

Recent tooling developments reflect a growing recognition of this need.
Positron~\cite{positron} extends the traditional IDE with a built-in data explorer
and native support for inspecting Pandas dataframes within a Python REPL or debugging session.
Nbdev~\cite{nbdev, howard2020} takes a different approach
by treating the notebook as the singular source of truth
from which production-ready Python modules,
tests,
and documentation are derived automatically.
Kedro~\cite{kedro} enforces software engineering best practices from the outset
by structuring ML projects around modular nodes and pipelines,
and extends this with interactive data lineage graphs,
dataset previews,
and experiment tracking through the Kedro-Viz plugin.
These tools represent a new generation of infrastructure
that seeks to close the gap between data science exploration and production engineering.
Although, empirical evaluation of their effectiveness in practice remain an open research priority.

A more disruptive development on the horizon is the rise of agentic coding systems,
where large language models plan, write, execute, and iteratively refine code
with minimal human supervision~\cite{watanabe2025}.
AI agents such as GitHub Copilot~\cite{copilot} and Claude Code~\cite{claudecode}
have been widely adopted across the software engineering community,
and are being used to develop new features, refactor code, update documentation and open pull requests~\cite{robbes2026}.

Can AI agents help reduce the debt incurred during prototyping,
by autonomously performing engineering changes currently performed by human practitioners?
Or will it further erode the developer's mental model of the program and accumulate \emph{cognitive debt}~\cite{storey2026generative, storey2026hearing,fowler2026fragments}?
These remain important and open questions for the SE4AI research community.

\subsection{Implications for Research}

Our findings provide
practitioner-grounded evidence for \textit{how}
the ML-specific technical debt identified by Sculley et al.~\cite{sculley2015}
manifests during the transition from prototype to production.
The data quality assertions (E10, Sec.~\ref{sec:testing}),
fail-safe checks (EC13, Sec.~\ref{sec:testing}),
and pipeline guardrails (EC18, Sec.~\ref{sec:pipeline}),
our participants described
are concrete instantiations of the monitoring and testing debt
that Sculley et al.\ warned about,
but which had not been empirically characterized at the level of individual engineering practices.
Similarly, the feedback loop mitigation strategies (EC8, Sec.~\ref{sec:datapipeline})
described by P6 provide concrete evidence for how practitioners address
the hidden feedback loops that Sculley et al.\ described abstractly.

Prior work on notebook quality~\cite{pimentel2019large-scale,
grotov2022large-scale, pimentel2021understanding}
used static analysis to identify deficiencies such as
low reproducibility rates,
poor coding practices,
and lack of modularity.
Our findings complement this work
by showing what practitioners \textit{do} about such quality problems
when the code must transition to production.
The engineering changes we identified can be understood as practitioners' responses
to the quality deficiencies that the notebook literature has documented.
Our findings show that the notebook quality problems
identified by Pimentel et al.\ and Grotov et al.\ are not merely academic concerns
but have concrete downstream costs in operationalization effort.

Our hybrid deductive-inductive approach for identifying quality attributes (RQ2)
provides an empirical bridge between the quality model
proposed by Siebert et al.~\cite{siebert2021construction}
and the concerns that ML practitioners face during operationalization.
Several attributes in the Siebert et al.\ framework,
such as maintainability, portability, and reliability,
were strongly represented in practitioner accounts,
suggesting that these attributes capture genuine concerns in practice.
At the same time,
our inductive analysis extended the framework
with attributes such as development velocity and cost efficiency,
that emerged from the specific context of operationalization.
The mapping between engineering changes (RQ1) and quality attributes (RQ2)
further demonstrates that quality concerns do not map one-to-one to engineering practices.
Individual changes often address multiple quality attributes simultaneously,
and the same quality concern may be addressed through different
engineering changes depending on organizational context and system constraints.

\subsection{Threats to Validity}

As with any qualitative research,
our findings should be understood as transferable rather than statistically generalizable.
We observed thematic saturation in the final two interviews, though we
acknowledge that saturation in qualitative research is
contested~\cite{aldiabat2018data, guest2006}
and that a larger sample might reveal additional
engineering changes or quality concerns,
particularly in domains not represented in our cohort.

Braun and Clarke~\cite{braun2006, braun2019} emphasize that codes in RTA reflect the
researcher's analytical interpretation of the data rather than inter-rater
agreement.
Methodological rigor was ensured through ongoing dialogue with all
authors, iterative refinement of codes and themes, and collaborative review of
emerging findings.

\section{Related Work}

\subsection{Empirical Studies of ML-Enabled Software Development}

A growing body of empirical research has examined the challenges of
building and deploying ML-enabled software systems. Lwakatare et
al.~\cite{lwakatare2020large-scale} conducted a systematic literature
review of 72 industry-focused papers and identified 23 challenges
across six domains, where adaptability and scalability emerged as the
most frequently reported concerns. Nahar et
al. (2023)~\cite{nahar2023meta-summary} extended this
landscape through a meta-summary of 50 prior studies, and cataloged
the challenges of building products with ML components.
At the team and organizational level, Nahar et
al. (2022)~\cite{nahar2022collaboration} interviewed 45 practitioners across
28 organizations and reported collaboration challenges that arise
when integrating ML components, such as difficulties in requirements
elicitation, data handoffs, and integration across team boundaries.
Their findings highlight how the exploratory nature of ML development
creates coordination friction.

Our study complements this work by
focusing on the engineering changes that are performed during operationalization, and ground these changes
in the software quality concerns that motivate them.

% Chen et al.~\cite{chencomprehensive2020} studied challenges in
% deploying deep learning software by analyzing 3,023 Stack Overflow
% posts and proposed a taxonomy of deployment challenges encountered by
% developers in practice. While their work identifies challenges through
% developer forum posts, our study collects primary data through
% semi-structured interviews, allowing us to examine not only what
% challenges practitioners encounter but how they resolve them through
% specific engineering changes.

% Sens et al.~\cite{sens2024large-scale} examined how ML models are
% integrated into ML-enabled systems by analyzing 2,928 open-source
% repositories, identifying integration patterns, reuse practices, and
% architectural characteristics. Their focus on system-level
% characteristics and artifact reuse complements our study's focus on
% the practitioner-level engineering work required to transition
% prototypes into those systems.

\subsection{Software Quality in ML Systems}

Prior works have characterized software quality concerns in ML systems
through prescriptive frameworks and architectural analyses.
Siebert et al.~\cite{siebert2021construction} proposed a quality model
grounded in the ISO/IEC 25010 standard~\cite{iso25002}.
Breck et al.~\cite{breck2017ml} proposed an ML test score rubric
enumerating quality requirements for production ML systems,
that cover infrastructure, monitoring, data, and model aspects.
Sculley et al.~\cite{sculley2015} identified
technical debt in ML systems at an architectural level,
such as monitoring debt, pipeline jungles, and feedback loop risks.
Cot\'{e} et al.~\cite{cote2024quality} examined software quality attributes
in data science workflows
and characterized the quality concerns
prevalent during the exploratory phase of ML development.

Our study complements this body of work
by providing evidence of how quality concerns manifest,
and are resolved when transitioning from prototype to production.
While Siebert et al. provides a structural framework
and Breck et al. prescribes what a mature ML system \textit{should} do,
our study
shows what practitioners \textit{actually} do,
and which quality concerns drive those decisions.
We adopt Siebert et al.'s model as the initial coding structure for our RQ2 analysis,
and extended it inductively with attributes that emerged from practitioner accounts.

Our findings further show how the technical debts described abstractly by
Sculley et al.\ manifest as concrete engineering changes during operationalization.
We extend the quality characterization by Cot\'{e} et al.\
from the exploratory phase into the operationalization phase,
by tracing how concerns shift
and new trade-offs emerge
as practitioners move from interactive notebooks to automated production systems.

\subsection{Code Quality in Computational Notebooks}

Pimentel et al.~\cite{pimentel2019large-scale} conducted a large-scale study
of 1.4 million Jupyter notebooks from GitHub,
and highlighted reproducibility challenges of the notebook format.
A follow-up study~\cite{pimentel2021understanding}
examined practices within notebooks that contribute to code quality
and found widespread violations of software engineering conventions such as
non-linear execution,
lack of version control,
and poor modularity.
Grotov et al.~\cite{grotov2022large-scale} confirmed these findings
by conducting a large-scale study to compare code quality
in Jupyter notebooks to Python scripts.

Our work connects these documented quality deficiencies to their downstream costs.
The engineering changes we identified can be understood as practitioners' responses
to the quality problems arising from the use of notebooks.

\section{Conclusion}

Our investigation of the engineering work involved
in transitioning ML prototypes from computational notebooks to production systems
through semi-structured interviews with 13 ML practitioners,
identified 23 engineering changes across five themes
and 20 software quality attributes that motivate them.
Our primary finding is that
prototyping in computational notebooks externalizes oversight to the human practitioner
and accumulates \textit{oversight debt}.
Practitioners repay this debt during operationalization
by constructing automated substitutes
for the interactive oversight that notebooks provide.
Unlike conventional technical debt,
oversight debt does not accumulate from poor engineering practices,
but from the notebook format itself.
Future work should examine whether these findings generalize across broader
geographic and organizational contexts,
and investigate how tooling and framework design can reduce the compensatory engineering burden.

\bibliographystyle{IEEEtran}
\bibliography{IEEEabrv,bibliography}

\end{document}